\documentclass[twocolumn,superscriptaddress,aps,prd,amsfonts,amsmath,amssymb,nofootinbib,longbibliography,preprintnumbers]{revtex4-1}

\usepackage[pdftex,colorlinks=true,plainpages=false,pdfpagelabels]{hyperref}
\usepackage[pdftex]{color}
\hypersetup{colorlinks = true}

\usepackage{xspace}
\newcommand{\cp}[1]{\ensuremath{\mathbb{CP}^#1}\xspace}
\newcommand{\cpk}{\cp{k}}

\newcommand{\kahler}{K\"ahler\xspace}

\begin{document}

\preprint{IPMU14-0335}

\title{Chaotic Inflation from Nonlinear Sigma Models in Supergravity}

\author{Simeon Hellerman}
\email{simeon.hellerman.1@gmail.com}
\affiliation{Kavli Institute for the Physics and Mathematics of the Universe (WPI)\\
             Todai Institutes for Advanced Study, The University of Tokyo\\
             Kashiwa, Chiba 277-8582, Japan}
\author{John Kehayias}
\email{john.kehayias@vanderbilt.edu}
\affiliation{Kavli Institute for the Physics and Mathematics of the Universe (WPI)\\
             Todai Institutes for Advanced Study, The University of Tokyo\\
             Kashiwa, Chiba 277-8582, Japan}
\affiliation{Department of Physics and Astronomy, Vanderbilt University\\
             Nashville, TN 37235, United States}
\author{Tsutomu T.~Yanagida}
\email{tsutomu.tyanagida@ipmu.jp}
\affiliation{Kavli Institute for the Physics and Mathematics of the Universe (WPI)\\
             Todai Institutes for Advanced Study, The University of Tokyo\\
             Kashiwa, Chiba 277-8582, Japan}

\begin{abstract}

  \noindent
  We present a common solution to the puzzles of the light Higgs or
  quark masses and the need for a shift symmetry and large field
  values in high scale chaotic inflation. One way to protect, for
  example, the Higgs from a large supersymmetric mass term is if it is
  the Nambu-Goldstone boson (NGB) of a nonlinear sigma model. However,
  it is well known that nonlinear sigma models (NLSMs) with nontrivial
  K\"ahler transformations are problematic to couple to
  supergravity. An additional field is necessary to make the K\"ahler
  potential of the NLSM invariant in supergravity. This field must
  have a shift symmetry --- making it a candidate for the inflaton (or
  axion). We give an explicit example of such a model for the coset
  space $SU(3)/SU(2) \times U(1)$, with the Higgs as the NGB,
  including breaking the inflaton's shift symmetry and producing a
  chaotic inflation potential. This construction can also be applied
  to other models, such as one based on
  $E_7/SO(10) \times U(1) \times U(1)$ which incorporates the first
  two generations of (light) quarks as the Nambu-Goldstone multiplets,
  and has an axion in addition to the inflaton. Along the way we
  clarify and connect previous work on understanding NLSMs in
  supergravity and the origin of the extra field (which is the
  inflaton here), including a connection to Witten-Bagger
  quantization. This framework has wide applications to model
  building; a light particle from a NLSM requires, in supergravity,
  exactly the structure for chaotic inflaton or an axion.

\end{abstract}

\maketitle

\section{Introduction and Motivation}

Over the past two years there have been several exciting experimental
results, which both confirm theories developed long before as well as
challenge us to better understand their origin. The discovery of the
Higgs boson~\cite{atlashiggs, *cmshiggs} brings renewed attention to
the issue of the apparent lightness of the Higgs mass compared to any
UV scale, like the Planck mass. The Higgs is not the only light field
we are puzzled over; the lightness (smallness of the Yukawa couplings)
of the first two generations of quarks is a longstanding
question. More recently, there has been much discussion on the
possible discovery of B-modes in the CMB by BICEP2~\cite{biceps}, but
which may be due to dust~\cite{planckdust} rather than primordial
gravitational waves. However, a large value for the tensor to scalar
ratio, $r \sim 0.1$, is still possible. 
Such a value, or more generally any motivations of models
for high scale or large field inflation like chaotic inflation
\cite{Linde:1983gd}, raises the question of how to control higher
dimensional operators which will not be suppressed in the inflaton
potential. In these models, where do such large field values (of order
or greater than the Planck scale) come from, and how are such models
consistent?

There are several ways to address these problems, although it is not
at all obvious that they could be closely related. Consider first the
Higgs mass, which can be generated by a supersymmetric mass term. One
requires some way to either generate a mass much smaller than the
supersymmetry scale, or else forbid this operator. If the Higgs is a
Nambu-Goldstone boson (NGB) of a $G/H$ nonlinear sigma model (NLSM)
\cite{cpkpaper}, this would do both of these things: a NGB is massless
at first approximation, and cannot have such a mass term. The Higgs
mass is then protected until we introduce operators which break
$G/H$. More generally, we can think of any light particle, such as the
first two generations of quarks, as a possible NGB (or fermion partner
under supersymmetry) from a NLSM.

However, as soon as we consider local supersymmetry, we run into well
known problems for coupling a NLSM to
supergravity~\cite{wittenbagger}. The reason is that the K\"ahler
potential has a nontrivial transformation,
\begin{equation}
  \label{eq:ktrans}
  K(\Phi, \Phi^\dag) \rightarrow K(\Phi, \Phi^\dag) + g(\Phi) + g^\dag(\Phi^\dag),
\end{equation}
with $g$ a holomorphic function. Such functions have no effect in
global supersymmetry when integrated over all of superspace, but in
local supersymmetry they do not disappear. Here, too, there are
several solutions which have been studied in the past~\cite{ks, ky}
(see also~\cite{Buchmuller:1983iu, *Kugo:1983ma}
and~\cite{GrootNibbelink:1998tz} for earlier work). Generally one must
consider a non-compact NLSM, $G'/H$, which can be coupled to
supergravity. The extension of the original compact manifold
necessarily contains a (at least one) new chiral superfield $Z$. It
may be surprising that this field must appear in the \kahler potential
as $Z + Z^\dag$, possessing a shift symmetry. In special cases, as in
Witten-Bagger models~\cite{wittenbagger}, the manifold can be compact
and does not require extra fields (instead there is the quantization
condition of the K\"ahler form), and we will discuss how these two cases may possibly be connected.

Thus we see we are led to a possible solution to our second problem of
how to forbid higher dimensional operators in the inflaton
potential. $Z$ will have a completely flat direction ($Z - Z^\dag$
does not appear in the potential) which is protected by the shift
symmetry. More generally, to embed chaotic inflation in supergravity
\cite{kyy1} one typically requires a shift symmetry, whose existence
is simply imposed. Here we have an immediate explanation for the
origin of this symmetry. In order to have an inflationary potential,
though, we will need to break this symmetry. To understand the large
initial value of the inflaton in chaotic inflation we will look at
Witten-Bagger models which naturally have large field values (larger
than the Planck mass). We can then construct phenomenological models,
which incorporate this solution to both the Higgs or light quark
masses and chaotic inflation problems, in detail.

This paper is organized as follows. In the following section,
Section~\ref{sec:nlsmsugra}, we will clarify the difficulties and
solutions to coupling a NLSM to supergravity. The key points have been
understood in the past, but a coherent picture is essential to this
work. Using a \cp1 model as our guide, we will relate various
proposals for coupling a NLSM to supergravity. Following that we will
look at an explicit model, one based on $SU(3)/SU(2) \times U(1)$
which includes the Higgs as a NGB. In Section~\ref{sec:shift} we
investigate how to break the shift symmetry and propose a connection
to Witten-Bagger models. Again we use \cp1 as the prototype before
applying to our model with the Higgs. Finally, in
Section~\ref{sec:discon} we will discuss the application to other
models, such as an $E_7/SO(10) \times U(1) \times U(1)$ model for two
generations of light quarks. This model includes an axion in addition
to the inflaton. Finally, we will comment on future directions.

\section{Nonlinear Sigma Models and Supergravity}
\label{sec:nlsmsugra}

Let us demonstrate the difficulties in coupling a NLSM to supergravity
by considering the basic $\cp1 \cong SU(2)/U(1)$ model. This model is
determined completely by its symmetries, which require a \kahler
potential of the Fubini-Study form,
\begin{equation}
  \label{eq:cp1k}
  K = f_\phi^2\log\left(1 + \frac{\phi\phi^*}{f_\phi^2}\right)
\end{equation}
where $\phi$ is the NG multiplet (a chiral superfield\footnote{In terms
  of the homogeneous coordinates of \cp1 $\phi$ is given by their
  ratio, an affine coordinate. The ``conjugate'' coordinate is the
  reciprocal.})  and $f_\phi$ is the scale of the effective theory
(i.e.~the decay constant). If we think of \cp1 as a manifold it is
$S^2$, the 2-sphere, and requires two coordinate patches. A K\"ahler
transformation takes us between these patches, with $\phi \rightarrow
f_\phi^2/\phi$. Then the K\"ahler potential transforms as $K(\phi, \phi^*) \rightarrow
K(f_\phi^2/\phi, f_\phi^2/\phi^*)$ and
\begin{align}
  \label{eq:cp1ktrans}
  K\left(\frac{f_\phi^2}{\phi}, \frac{f_\phi^2}{\phi^*}\right) &= f_\phi^2\log\left(1 + \frac{f_\phi^2}{\phi\phi^*}\right)\nonumber\\
  = K(&\phi, \phi^*) - f_\phi^2\log\left(\frac{\phi}{f_\phi}\right) - f_\phi^2\log\left(\frac{\phi^*}{f_\phi}\right).
\end{align}
Indeed, we see we have the form $K \rightarrow K + g + g^*$ with $g$ a
holomorphic function. In global supersymmetry then $g$ drops out when
integrating $\int\mathrm{d}^4\theta K$ in the Lagrangian. In local
supersymmetry, however, these terms remain and we must think more
carefully of how to couple to supergravity.

We can make $K$ invariant with an additional field, $Z$, with the
right transformation properties~\cite{ks, ky}. The \kahler potential
is then
\begin{equation}
  \label{eq:ksk}
  K = K(\phi, \phi^*) + f_\phi(Z + Z^*),
\end{equation}
where we have used $f_\phi$ for dimensional reasons (the only other scale
is the Planck mass, but we will see later these are related). This
\kahler potential is incomplete, as $Z$ does not have a kinetic term
without considering higher order terms. We will consider such
additional terms below. Under a \kahler transformation we must have
\begin{equation}
  \label{eq:ztrans}
  Z \rightarrow Z + f_\phi\log\left(\frac{\phi}{f_\phi}\right),
\end{equation}
and similarly for $Z^*$. We see that $Z$ possesses a shift symmetry
and the imaginary part will have no potential with such a symmetry (to
build an inflationary model we will of course need to break the shift
symmetry).

Now we make an interesting observation: $Z$ \textit{must be charged}
under the (assumed) linearly realized $U(1)$ of \cp1. Thus $SU(2)$ is
actually completely broken and the additional flat direction is a
consequence of this breaking. This can easily be seen by using the
Jacobi identity for the $SU(2)$ generators on the field $Z$. In order
for it to be satisfied, $Z$ must transform under the $U(1)$. This
requirement for breaking the $U(1)$ is another understanding for how
to couple a NLSM to supergravity~\cite{ky}.

The origin of $Z$ can also be seen from considering how a nonlinear
sigma model comes from a linear sigma model (which we can
straightforwardly couple to supergravity) in supersymmetry. It has
been long known that supersymmetric NLSMs, with a linear sigma model
origin, must come with additional degrees of freedom, called
quasi-NGBs \cite{Buchmuller:1983iu, *Kugo:1983ma}. Generically, these
quasi-NGBs double the NGB degrees of freedom. However, if one has a
special \kahler potential (non-generic), it is possible to realize the
NLSM with fewer quasi-NGBs. We will show such a specific model
below. With this interpretation, the $Z$ field has a clear origin and
reason for its associated flat direction.

\subsection{A Model for a Light Higgs and Chaotic Inflation}
\label{sec:model}
As we discussed above, one possible resolution to the puzzle of the
lightness of the Higgs mass is if the Higgs is a NGB of some NLSM. In
fact, we have previously constructed such a model in~\cite{cpkpaper}
(which is an extension of the work in~\cite{cp1paper}). In this model
the Higgs is the NGB of a $SU(3)/SU(2)_L \times U(1)_Y$ NLSM, as it
has exactly the right quantum numbers. 

However, we know we cannot simply couple such a NLSM to supergravity,
as we have seen above. Since the extra field in the construction
of~\cite{ks} must break the $U(1)$ (which is the requirement in the
language of~\cite{ky}), the group structure is actually
$SU(3)/SU(2)$. This is equivalent to a $U(3)/SU(2) \times U(1)$ NLSM
which has a known construction~\cite{Kugo:1984tj, *Goto:1990me} (see
also~\cite{ky}) with an invariant $K$ and a quasi-NGB. We identify the
unbroken $SU(2)$ as the weak gauge group of the Standard Model, and
the NG superfield is one (the lightest) of the Higgs multiplets, $H_u$
or $H_d$.

The NG superfield is an $SU(2)$ doublet, labeled $(\phi_1, \phi_2)$, and
the quasi-NGB chiral superfield is $Z$ (which has been dubbed the
``novino'' in~\cite{Buchmuller:1983na}). Defining the matrix $\xi$ as
\begin{equation}
  \xi \equiv \left[
  \begin{matrix}
    e^{\kappa Z} & 0\\
    0      & e^{\kappa Z}\\
    \phi_1     & \phi_2
  \end{matrix}
  \right],
\end{equation}
where $\kappa$ has mass dimension $-1$ and is related to the scale of the
NLSM, the K\"ahler potential can be written as~\cite{Kugo:1984tj, *Goto:1990me}
\begin{equation}
  \label{eq:u3kahler}
  K = -F(\det\xi^\dag\xi),
\end{equation}
for a function $F$, subject to constraints for proper kinetic terms
and vacuum.

Under the global $U(3)$ transformation the matrix $\xi$ transforms as
\begin{equation}
  \xi \rightarrow g\xi h^{-1},
\end{equation}
where $g \in U(3)$ and $h \in SU(2)$. Thus $\det(\xi^\dag\xi)$, and therefore
$K$, is invariant under the global $U(3)$ transformations. There is no
difficulty then in coupling this model to supergravity.

We can connect directly to the work of~\cite{ks} and the \kahler
potential of the form in eq.~\eqref{eq:ksk} by the following field
redefinitions. First, we define and write explicitly
\begin{equation}
  \label{eq:u3xdef}
  x \equiv \det\xi^\dag\xi = e^{2\kappa(Z + Z^\dag)} + e^{\kappa(Z + Z^\dag)}\left(|\phi_1|^2 + |\phi_2|^2\right),
\end{equation}
which can be rewritten as
\begin{equation}
  \label{eq:u3xrewrite}
  x = e^{2\kappa(Z + Z^\dag)}\left(1 + \phi_i'\phi_i'^\dag \right),
\end{equation}
with the field redefinition $\phi' = e^{-\kappa Z}\phi$. Now define
\begin{equation}
  \label{eq:ksydef}
  y \equiv \log x = \log(1 + \phi_i'\phi_i'^\dag) + 2\kappa(Z + Z^*),
\end{equation}
which is exactly the field combination in~\cite{ks} with \kahler
potential
\begin{equation}
  \label{eq:kskahler}
  K\left( \log(1 + \phi_i'\phi_i'^\dag) + 2\kappa(Z + Z^*) \right) = K(y).
\end{equation}
Thus we have an exact equivalence between the two different looking
models. The group structure is equivalent once one notices that the
$U(1)$ is actually broken in~\cite{ks}, and the counting of flat
directions is the same. It is clear that $Z$ has a shift symmetry and
can play the role of the inflaton. While $Z$ comes from an extended
supergravity multiplet in~\cite{ks}, we see that it is equivalent to a
quasi-NGB origin as in~\cite{Kugo:1984tj, *Goto:1990me, ky}

\section{Large Field Values and Shift Symmetry Breaking}
\label{sec:shift}

While we now have a model demonstrating how a NLSM in supergravity has
the right starting point for inflation, there are two key
questions. How do we break the shift symmetry and produce an
inflationary potential, and how do we achieve the large field values
necessary for chaotic inflation? We will use a simpler \cp1 model to
answer these questions, and the extension is straightforward to the
$SU(3)/SU(2)$ model above, or other \cpk models.

To successfully have inflation we must break the shift symmetry of $Z$
and generate a potential for its imaginary part. Perhaps the simplest
way to accomplish this is to add a superpotential with a new field,
$X$, and coupling
\begin{equation}
  \label{eq:u3w}
  W = mXZ.
\end{equation}
We need $m \sim 10^{-5}M_p$ for chaotic inflation~\cite{kyy1, kyy2} (see
also the recent work of~\cite{Harigaya:2014roa}). This breaking is
technically natural as we have the shift symmetry as we take $m
\rightarrow 0$.\footnote{In the $SU(3)/SU(2)$ model a term linear in
  $X$ in the superpotential, $cX$, is allowed by all
  symmetries. Shifting $Z$ to cancel this term normally leads to a
  large linear term in $K$ which must be small or inflation will not
  end~\cite{kyy2}. By viewing the model written in terms of $x$ we see
  this is not a problem: a real shift in $Z$ is equivalent to an
  overall factor and can be absorbed, while any imaginary part cancels.}

There are several difficulties in trying to introduce such a
superpotential in a NLSM in supergravity. Besides breaking the shift
symmetry of $Z$, this superpotential will break \kahler invariance and
$G$-invariance due to the properties of $Z$ (which were required to
couple to supergravity in the first place). However, $G$ is just a
global, nonlinearly realized symmetry and hence a violation is not a
serious problem (and motivates such a small value for $m$).

A more serious problem is that we need to have very large (in Planck
units) field values for chaotic inflation, yet we naturally expect
$f_\phi < M_p$, with $M_p$ the reduced Planck mass. Therefore it seems
desirable to have values of $f_\phi$ larger than $M_p$. One of the few
such models is that of Witten and Bagger~\cite{wittenbagger}. Thus we
wish to see if there can be a connection to Witten-Bagger theories as
a special case of the models we are studying to explain large field
values.

Consider if $Z$ is just a chiral superfield, invariant under \kahler
and $G$ transformations, but with a shift symmetry. In this case, to
be coupled to supergravity, the \cp1 NLSM would have to be a
Witten-Bagger theory, with the decay constant constrained to be
quantized in units of $M_p$, and thus larger than $M_p$. One gains the
benefit of naturally explaining large field values (we expect fields
to take values of order the decay constant), but pays the price of
losing an origin for the shift symmetry and all interactions in the
superpotential (we must have $W = 0$ in a Witten-Bagger model).

We propose a connection between these different types of theories, one
where $Z$ and its shift symmetry are put in by hand and another where
$Z$ is required by supergravity, by the following \kahler potential
with real, dimensionless parameters $a, b$:
\begin{align}
  \label{eq:conj}
  K = &f_\phi^2\left[\log\left(1 + \frac{\phi\phi^*}{f_\phi^2}\right) + \frac{1}{f_\phi}(Z + Z^*)\right]\nonumber\\
      &+ \frac{a^2}{2}f_\phi^2\left[\log\left(1 + \frac{\phi\phi^*}{f_\phi^2}\right) + \frac{1}{f_\phi}(Z + Z^*)\right]^2\nonumber\\
      &+ \frac{b}{2}(Z + Z^*)^2 + XX^*.
\end{align}
The superpotential is only turned on when it is allowed, namely when
$a \neq 0,~b = 0$ and the theory is in the
Komargodski-Seiberg-Kugo-Yanagida (KSKY) branch. In general, the first
two terms could be replaced with some general function of the \kahler-
and $G$-invariant quantity in square brackets (subject to appropriate
constraints for canonical kinetic terms, etc.). There may also be
higher order terms besides what is written above, but these are not
relevant for this discussion. When $a = 0, b \neq 0$, we are in a
Witten-Bagger theory and the term with coefficient $b$ is a kinetic
term for $Z$. We conjecture that the KSKY branch is smoothly connected
to the Witten-Bagger branch through the parameters $a$ and $b$. This
implies that $f_\phi^2 = 2nM_p^2$, for some integer $n$, everywhere.

We now consider the theory on the KSKY branch as a model for chaotic
inflation. From the \kahler potential above with $b = 0$, we can
canonically normalize the kinetic term for $Z$ by the field
redefinition $aZ \rightarrow Z$. Then the linear term in $Z$ has the
dimensionless combination $(Z + Z^*)/af_\phi$, and thus we expect the
initial value of $Z$ to be of order $af_\phi$. For chaotic inflation
we require this combination to be $\mathcal{O}(10)$ in Planck
units. Furthermore, this linear term has overall coefficient
$f_\phi/a$, which must be less than order one for inflation to
end~\cite{kyy1}. Thus we need that $f_\phi \sim a$ and with $f_\phi
\sim \sqrt{n}$ by Witten-Bagger quantization (in Planck units, and $n
\in \mathbb{Z}$), we have that $a \sim \sqrt{n}$. Finally, we see that
we only need $n \sim 10$ to satisfy $af_\phi \sim \mathcal{O}(10)M_p$
which is very reasonable.

We have now the following picture. The additional field $Z$ comes from
considering the general case of coupling to supergravity when $f_\phi$
is not quantized. This is the origin of the inflaton and its shift
symmetry (in the limit $m \rightarrow 0$). We then flow to the
Witten-Bagger theory: $f_\phi$ must be quantized and larger than
$M_p$, explaining the necessary large field values. Here $Z$ is just a
field with a shift symmetry and does not transform in any special way
to couple the theory to supergravity.

\section{Discussion and Conclusion}
\label{sec:discon}

We can easily apply the above work to other NLSMs. Consider the
$E_7/SO(10) \times U(1) \times U(1)$ NLSM~\cite{Kugo:1983ai,
  *Yanagida:1985jc, *Evans:2013uza} which contains two $16$ NG
multiplets which can be identified with the light two generations of
quarks and leptons. It is interesting that their small Yukawa
couplings can be explained in this model as a weak breaking effect. To
couple this model to supergravity the two $U(1)$s should be
broken~\cite{ky}. Again, we can identify $Z$ with the inflaton, but we
have an additional NG superfield, $Z'$. It is tempting to identify
this with the QCD axion multiplet in order to solve the strong CP
problem\footnote{We have to assume that the mass term for the axion
  generated by explicit breaking of the global $E_7$ is negligible.}.

If we use the \kahler manifold $G/H = E_7/SU(5) \times U(1)^3$, we
have three families of quarks and leptons as the NG multiplets. In
this case we should introduce three singlets, $Z,~Z',$ and $Z''$ to
couple to supergravity. The $E_7$ manifold has three submanifolds,
$E_7/E_6 \times U(1), E_6/SO(10) \times U(1),$ and $SO(10)/SU(5)
\times U(1)$. When we intorduce an explicit breaking for $E_7
\rightarrow E_6$, the third family of quarks and leptons will have
Yukawa couplings and the third singlet, $Z''$, gets a corresponding
mass, and so on for the second generation. The very small Yukawa
coupling of the up quark, $\sim 10^{-5}$, would be due to the good
$SO(10)$ symmetry. Thus, when we introduce an explicit breaking of
$SO(10)$ the up quark has a Yukawa coupling and the singlet $Z$ gets a
mass. The lightest singlet $Z$ is the inflaton. Since the Yukawa
coupling for the up quark and inflaton mass arise from the $SO(10)$
breaking, it is plausible to have a relation~\cite{desytalk} between
them,
\begin{equation}
  \label{eq:so10relation}
  Y_\mathrm{up} \simeq m_\mathrm{inflaton}/M_p.
\end{equation}
It is perhaps surprising that this relation is almost satisfied for
chaotic inflation with $m_\mathrm{inflaton} \simeq 10^{13}$ GeV.

It is straightforward to apply the above techniques to other models as
well. A NLSM explanation for a light field requires other flat
directions, suitable for inflation or axions, once coupled to
supergravity. Our phenomenological model which may connect the present
KSKY branch to the Witten-Bagger branch may be realized in some higher
dimensional theories.  The embedding of the
Bagger-Witten realization of SUGRA into an ultraviolet-complete theory
has been little explored, and we hope that our model
may motivate further research in this direction.
However, this is beyond the scope of the present
paper.

\begin{acknowledgments}
  \noindent
  This work was supported by the World Premier International Research
  Center Initiative (WPI Initiative), MEXT, Japan. The work of
  S.H.~was also supported in part by a Grant-in-Aid for Scientific
  Research (26400242) from the Japan Society for Promotion of Science
  (JSPS). J.K.~was supported in part by the DOE (DE-SC0011981). This
  work was also supported by Grant-in-Aid for Scientific research from
  the Ministry of Education, Science, Sports, and Culture (MEXT),
  Japan, No.~26104009 and 26287039 (T.T.Y.).
\end{acknowledgments}

\bibliography{chaoticinflation_nlsm_refs}

\end{document}